# Stability of Inviscid Parallel Flows between Two Parallel Walls

**Hua-Shu Dou**

Temasek Laboratories,
National University of Singapore,
Singapore 117411
Email: tsldh@nus.edu.sg; huashudou@yahoo.com

**Abstract**: In this paper, the stability of inviscid parallel flow between two parallel walls is studied. Firstly, it is obtained that the velocity profile of the base flow for this classical problem is a uniform flow. Secondly, it is shown that the solution of the disturbance equation is $c_r = U$ and $c_i = 0$, i.e., the propagation speed of the disturbance equals the flow velocity and the disturbance in this flow is neutral. Finally, it is suggested that the classical Rayleigh theorem on inflectional velocity instability is incorrect which states that the necessary condition for instability of inviscid parallel flow is the existence of an inflection point on the velocity profile.

**Keywords**: Flow instability; Inviscid parallel flow; Rayleigh theorem; Inflection.

# 1. Introduction

Stability of parallel flows is the basis of modern flow stability theory. The Rayleigh theorem on inflectional instability is a fundamental theorem on inviscid stability theory. This theorem is found in many text books and employed in the scientific community since Rayleigh published his classical work in 1880 [1]. However, this theorem is facing some challenges today as it appears to contradict some experimental observations.

In the classical theory for flow instability, Rayleigh (1880) first developed a general linear stability theory for inviscid parallel shear flows, and showed that a necessary condition for instability is that the velocity profile has a point of inflection [1]. Heisenberg (1924) showed that if a velocity distribution allows an inviscid neutral disturbance with finite wave-length and non-vanishing phase velocity, the disturbance with the same wave-length is unstable in the real fluid when the Reynolds number is sufficiently large [2]. Later, Tollmien (1935) succeeded in showing that Rayleigh's criterion also constitutes a sufficient condition for the amplification of disturbances for velocity distributions of the symmetrical type or of the boundary-layer type [3]. However, Lin (1944) has shown where a point of inflexion is present in the velocity distribution, a neutral disturbance does not exist [4]. In other word, Rayleigh's criterion is not a sufficient condition for instability (the reason was further clarified by Fjϕrtoft (1950) [5]). In the analysis for boundary layer flow, it is found that viscosity may play dual roes with respect to stability. Lin was able to demonstrate the different influences of viscosity on the disturbance amplification at low Re and high Re. His conclusions are as follows. For small viscosity (high Re), the effect of viscosity is essentially destabilizing and an increase of Re gives rise to more stability. For large viscosity (low Re), viscosity plays a stabilizing role via the dissipation of energy. Fjϕrtoft (1950) gave a further necessary condition for inviscid instability, that there is a maximum of vorticity for instability; he also gave the second further necessary condition for inviscid instability, $U''(U - U_{IP}) < 0$ (see Fig.1) [5]. Therefore, it is well known that inviscid flow with inflectional velocity profile is unstable, while inviscid flow with no inflectional velocity profile is stable [6-10].

**(1) Rayleigh theorem contradicts to experimental observations**. The pipe Posiseuille flow and plane Couette flow are judged as inviscid stable according to Rayleigh theorem since there is no inflection point on their velocity profiles. This means that these flows for real fluids are stable in the inviscid limit. However, there is no doubt that these flows for real fluid are actually unstable and transit to turbulence at certain high Re. This may indicate that viscosity induces flow instability. On the other hand, from experiments [11-12], one can deduce that viscosity always plays stable role in these flows. Thus, *Rayleigh theorem appears to contradict the experimental observations in these flows*. A possible explanation is that these velocity profiles without inflection point are really inviscid unstable, and viscosity addition makes the flow becoming stable. When Re tends towards infinity, the flow becomes unstable, which is consistent with the assumed inviscid instability. Such explanation seems more reasonable. On consideration of these experimental observations, the Rayleigh theorem needs to be further examined.

**(2) Rayleigh theorem contradicts to energy gradient theory.** Recently, we proposed a new approach, named as *energy gradient theory (EGT)*, to explain flow instability and transition to turbulence [13-18]. The critical condition calculated at turbulent transition as determined by experiments shows consistent agreement among parallel flows [13-14] and Taylor-Couette flows [15]. When the theory is considered for both parallel and curved shear flows, three important theorems on flow stability have been deduced [17]. These theorems are: (1) Potential flow (inviscid and irrotational) is stable. This theorem is easily understood since a potential flow is of a uniform energy filed. A disturbance could not be amplified in a uniform energy filed. (2) Inviscid rotational (inviscid and nonzero vorticity) flow is unstable. This theorem means that inviscid rotational flow is unstable and the addition of viscosity may make the flow stable, which is consistent with the experimental observations. This theorem also means that *an inflection point on the velocity profile in inviscid flow is not a necessary condition for instability*, which contradicts the Rayleigh theorem. (3) Velocity profile with an inflectional point is unstable for pressure driven flows, for both inviscid and viscous flows. This theorem is in accord with several experimental observations and numerical simulations for viscous flows [11,12,18-22]. The energy gradient theory also showed that viscosity has only stabilizing role in parallel flows and circular flows [17-18].

Since the Rayleigh theorem and the energy gradient theory contradict each other, there is at least one to be wrong between them. Following these comparisons and analyses, there is the motivation to examine further the validity of the Rayleigh theorem. In present study, the stability problem of inviscid parallel flow between two parallel walls is employed for the study. It will be shown that the classical Rayleigh Theorem on inflectional velocity instability is incorrect.

## 2. Rayleigh Equation

It should be distinguished between the base flow and the mean flow. However, for linear disturbance, since the perturbation is infinitesmall, the mean flow is the same as the base flow, but, the concept is different. In the following, we employ the formulation found in [7].

Let the base flow, which may be regarded as steady, be described by its Cartesian velocity components U,V,W and its pressure P, the corresponding quantities for the disturbance will be denoted by u', v', w' (u' in streamwise, v' in transverse, and w' in spanwise directions) and p', respectively. Hence, in the resultant motion the velocity components and the pressure are

$$u = U + u', \ v = V + v', \ w = W + w', \ p = P + p'. \qquad (1)$$

Substituting the above expressions into the Euler equation for inviscid flow and subtracting the equation for the base flow, the linearized equation of disturbance can be obtained [1, 6-10].

It is assumed that the disturbance is two-dimensional (2D), then a stream function is introduced. The stream function representing a single oscillation of the disturbance is assumed to be of the form

$$\psi(x, y, t) = \phi(y) e^{i(\alpha x - \beta t)}, \qquad (2)$$

where $\alpha$ is a real quantity and $\beta$ is a complex quantity, $\beta = \beta_r + i\beta_i$. Dividing $\beta$ by $\alpha$, a complex quantity c is obtained, $c = \beta/\alpha = c_r + ic_i$. Here, $c_r$ is the speed of the wave propagating and $c_i$ expresses the degree of damping or amplification of the disturbance ($c_i = 0$, neutral disturbance; $c_i < 0$, the disturbance decays; $c_i > 0$, the disturbance amplified). Thus,

$$u' = \frac{\partial \psi}{\partial y} = \phi'(y)e^{i(\alpha x - \beta t)}, \tag{3}$$

$$v' = -\frac{\partial \psi}{\partial x} = -i\alpha\phi(y)e^{i(\alpha x - \beta t)}. \tag{4}$$

Introducing these values into the linearized inviscid equation of the disturbance, the following ordinary differential equation is obtained [1, 6-10],

$$(U - c)(\phi'' - \alpha^2 \phi) - U''\phi = 0, \tag{5}$$

which is known as the frictionless stability equation, or Rayleigh's equation. For inviscid parallel flow between two walls,

$$y = y_1, y_2, \quad \phi = 0. \tag{5a}$$

## 3. Rayleigh's necessary condition for instability of inviscid flows

Following the formulation in [7], one can re-write Eq.(5) as

$$\phi'' - \alpha^2 \phi - \frac{U''}{U - c}\phi = 0. \tag{6}$$

Next, we multiply Eq.(6) by $\phi^*$, the complex conjugate of $\phi$, then obtain [1, 6-10]

$$\phi^* \phi'' - \alpha^2 \phi \phi^* - \frac{U''}{U - c}\phi \phi^* = 0. \tag{7}$$

Then, integrating the above equation by part over $y$, using Eq.(5a), the imaginary part of the resulting equation is

$$c_i \int_{y_1}^{y_2} \frac{U''|\phi|^2}{|U - c|^2} dy = 0. \tag{8}$$

If the disturbance is amplified, $c_i$ is larger than zero. It can be seen that for the equality to be valid $U''$ has to change sign over the integration space. Thus, there should be at least one point over the distance between $y_1$ and $y_2$ at which $U''=0$. In other words, *it is necessary that there is an inflection point on the velocity profile for flow instability.* This is the famous Rayleigh theorem **[1].**

**4. Re-visiting: Solution of Rayleigh Equation**

**4.1 Solution of the Base Flow**

Before the stability of a linear disturbance to a base flow is analyzed, the base flow should be first solved. This can be done for the following inviscid parallel flows.

In the Cartesian coordinates expressed in Fig.2, the Euler equation and the continuity equation for steady incompressible flow read,

$$U\frac{\partial U}{\partial x} + V\frac{\partial U}{\partial y} = -\frac{\partial p}{\partial x}, \qquad (9)$$

$$U\frac{\partial V}{\partial x} + V\frac{\partial V}{\partial y} = -\frac{\partial p}{\partial y}, \qquad (10)$$

$$\frac{\partial U}{\partial x} + \frac{\partial V}{\partial y} = 0. \qquad (11)$$

The boundary conditions are

$U \neq 0$ and $V = 0$ at $y = \pm h$.

As is well known, any shapes of velocity profile of $U$ **satisfy** the above Euler equations and the boundary conditions. However, not all these velocity profiles are the physical **solutions** of the Euler equations. Actually, there is a unique solution of the Euler equation between two parallel walls which accords with the physics. This unique solution is U(y)=Const. In the following, we will give a complete and strict proof on the uniqueness of the solution of the Euler equations within two parallel walls.

In above Eqs.(9-11), there are three unknowns *U, V* and *p* with the three equations, and thus the system is closed.

Since $V \equiv 0$ for the parallel flow, then we have $\frac{\partial p}{\partial y} = 0$ from Eq.(10).

Since $V \equiv 0$ for the parallel flow, then we have $\frac{\partial U}{\partial x} = 0$, or $U = U(y)$ from Eq.(11). With these, we obtain $\frac{\partial p}{\partial x} = 0$ from Eq.(9).

Considering $\frac{\partial p}{\partial x} = 0$ and $\frac{\partial p}{\partial y} = 0$, we obtain,

$$p(x,y)=\text{Const.} \tag{12}$$

Thus, the pressure p is a constant in inviscid parallel flows.

Introducing Eqs.(9) and (10) into Eq.(11), we obtain the functional relation,

$$f(U,V,P) = 0. \tag{13}$$

Now, we introduce a ***constraint*** $V \equiv 0$ for parallel flows to be considered, Eq.(13) becomes

$$f(U,P) = 0, \tag{14}$$

or $\quad$ U=f(p). $\tag{15}$

Thus, U=U(y) is a function of p only. Since p is a constant in the flow field (Eq.(12)), we have,

$\quad$ U(y)=Const. $\tag{16}$

The constant in Eq.(16) is determined by the slip velocity at the walls.

Thus, the only available solution of Euler equation between two parallel walls is the uniform flow (see Fig.2).

Alternatively, above conclusion can also be directly obtained as follow. Let us set $V \equiv 0$ for the problem of the inviscid parallel flow in Eqs.(9) to (11), then the Eq. (10) is removed or Eq.(10) does not exist. Thus, the Eqs.(9) and (11) become Eqs.(17) and (18), respectively,

$$U \frac{\partial U}{\partial x} = -\frac{\partial p}{\partial x}, \tag{17}$$

$$\frac{\partial U}{\partial x} = 0. \tag{18}$$

The boundary conditions are

$$U \neq 0 \quad \text{at} \quad y = \pm h.$$

The Eqs.(17) and (18) show that the inviscid parallel flow becomes a pure one-dimensional (1D) flow problem. There is no any U dependent on y. This means that all the U along y direction is the same (**slug flow**). Obviously, the solutions for Eq.(17) and (18) are respectively as,

$$p(x) = cons \tan t, \tag{19}$$

$$U(x) = cons \tan t. \tag{20}$$

Therefore, both the velocity U and the pressure p are constants in the whole flow field. The solutions of Eqs.(19) and (20) are consistent with Eqs.(12) and (16). In summary, it is concluded that **the unique solution of the Euler equations between two parallel walls is the uniform flow (slug flow)** (see Fig.2).

As pointed out before, we know that there are numerous of arbitrary solutions for the Euler equation for parallel flows. We need to explain why these arbitrary solutions of the Euler equation for parallel flows are incorrect physically. In the flow field, all the flow variables are coupled, and they are mutually dependent each other. In other words, the contributions from these variables are balanced each other. For parallel flows, these variables are the velocity U and the pressure p since the velocity V is zero everywhere. Thus, the distribution of U(y) is balanced by the variation of the pressure p. If U(y) is non-uniform along the y direction, the pressure p should be non-uniform too. This contradicts the observations in Eqs.(9-11) that $\frac{\partial p}{\partial x} = 0$ and $\frac{\partial p}{\partial y} = 0$. As such, in order to accordance to the flow physics ($\frac{\partial p}{\partial x} = 0$ and $\frac{\partial p}{\partial y} = 0$), U(y) should be uniform along the y direction. Therefore, it is concluded that **the arbitrary solution of the Euler equation for parallel flows is incorrect physically.**

**4.2 Solution of the Disturbance Equation**

The solution of the disturbance equation Eq.(5) is shown below with the unique solution of the base flow between two parallel walls. From Eq.(16), we obtain

$$U' = 0; \quad and \quad U'' = 0. \tag{21}$$

Introducing $U'' = 0$ into Eq.(5), we have

$$(U - c)(\phi'' - \alpha^2 \phi) = 0. \tag{22}$$

There are two possible solutions for Eq.(22),

$$(U - c) = 0, \tag{23}$$

and

$$(\phi''-\alpha^2\phi) = 0. \tag{24}$$

For Eq.(23), we have the solution (noticing $c = c_r + ic_i$),

$$c_r = U = C \quad \text{and} \quad c_i = 0. \tag{25}$$

For Eq.(24), we have $\phi'' = \alpha^2\phi$, and the general solution to this differential equation is then

$$\phi(y) = B_1 e^{\alpha y} + B_2 e^{-\alpha y} \quad (B_1 \text{ and } B_2 \text{ are constants}). \tag{26}$$

Applying the boundary condition, $y = \pm h$, $\phi = 0$ (see Fig.2); then, it is found that Eq.(26) has no solution except $\phi(y) \equiv 0$. Thus, it is concluded that Eq.(24) has no solution under the given boundary conditions.

Therefore, the linear perturbation equation of inviscid flow between two parallel walls has only one solution $c_r = U = C$ and $c_i = 0$. This means that the propagating speed of the disturbance in inviscid uniform flows is a constant and the amplification rate of the amplitude of the disturbance is neutral.

As can be seen above, since the base flow is $U = C$ with $U'' \equiv 0$ in the whole domain and Eq.(5) degrades to Eq.(22), Eq.(6) needs not necessarily exist for inviscid parallel flows. Thus, the Rayleigh criterion may not exist. As such, the Rayleigh theorem on the inflectional instability of inviscid parallel flow is incorrect.

It is seen from Eqs.(12) and (16) or Eqs.(19) and (20) that *the uniform flow is the unique solution for inviscid parallel flow which satisfies the Euler equation and the slip boundary conditions*. Thus, it is impossible that there are options of inflection point existing or not on the velocity profile of the base flow for the inviscid parallel flow.

The reason why Rayleigh theorem is incorrect is due to that the solution of the base flow is assumed to have arbitrary shapes of velocity profiles (as in Fig.1) which is incorrect physically. In present study, **we introduce the constraint $V \equiv 0$ for inviscid parallel flows**, a unique solution for the base flow is obtained, i.e., uniform flow. Since the constraint $V \equiv 0$ is truly suitable for parallel flows, this solution accords with the physics.

## 5. Discussion

The energy gradient method showed that *the necessary and sufficient condition for instability of inviscid parallel flows is the non-zero vorticity* [17]. According to this criterion, for pipe Poiseuille flow, plane Poiseuille flow, and plane Couette flow, they are inviscid unstable. The addition of viscosity causes the flow to become less unstable. When the Re tends to infinity, the flow approaches the inviscid flow limit and is hence unstable in terms of criterion. Therefore, this criterion is consistent with the experimental observations [11,12,19,20].

Figure 3 shows the evolution of the stability bahaviour of velocity profile between two parallel walls with the variation of viscosity $\mu$. It can be observed:

(a) When $\mu$ is large (Re is low), the laminar profile of parabola between the two parallel walls is stable.

(b) When $\mu$ is reduced (Re is high), the velocity profile becomes less stable.

(c) When $\mu$ tends to zero (Re tends to infinite), the flow approaches the inviscid flow state; it is unstable according to the energy gradient method.

(d) When $\mu$ is zero (inviscid flow), the velocity profile between the two parallel walls is a uniform flow. It is in a stable state of inviscid flow.

If the velocity profile of parabola between the two parallel walls is assumed to be inviscid (e.g., Fig.3(c)), this profile is unstable due to the non-zero vorticity, and it is not in a stable state. It will attempt to approach its stable state: uniform flow (Fig.3(d)). The role of inviscid flow part in real flows is always trying to flatten the velocity profile tending towards the uniform flow. In most cases, the flow is always trying to assume its most stable state under the roles of *energy gradient* and perturbation. This is the reason why instability occurs in inviscid flow as well as in viscous flow.

The existence of inflection point on the velocity profile is related to instability of viscous flows [18]. The energy gradient method showed that an inflection point on the velocity profile leads to instability for transition to turbulence for pressure driven flows [18], but for shear-driven flows, its existence is neither a necessary nor a sufficient condition for turbulent transition [19].

## 5. Conclusions

The inviscid base flow between two parallel walls is the uniform flow from the Euler equation with the slip boundary condition. Its first and second derivatives are both zero everywhere. For this flow, the solution of the linearly perturbed equation (Rayleigh equation) is $c_r = U$ and $c_i = 0$. That is, the propagating speed of disturbance is the same as that of base flow and the disturbance is neutral. There is no possibility of inflection point existence in inviscid parallel flows. As a result, the inviscid parallel flow between two parallel walls is stable and it is suggested that the classical Rayleigh theorem is incorrect.

The energy gradient method showed that the necessary and sufficient condition for instability of inviscid parallel flows is the non-zero vorticity. This theorem is consistent with the experimental observations. The energy gradient method also showed that velocity profile with an inflectional point is unstable for pressure driven flows for viscous fluids. This theorem is in accord with several experimental observations and numerical simulations for viscous flows.

**Acknowledgement**

I thank Prof. BC Khoo from National University of Singapore and Prof. ZS She from Peking University for their helpful comments.

**References**

1  L. Rayleigh, On the stability or instability of certain fluid motions. Proc. Lond. Maths. Soc., 1880, 11: 57-70
2  W. Heisenberg, Uber Stabilitat und Turbulenz von Flussigkeitsstromen, Ann Phys., Lpz. (4) 74, 1924, 577-627, On stability and turbulence of fluid flows, NACA TM-1291, 1951.
3  W. Tollmien, 1935, Ein allgemeines Kriterium der Instabilitat laminarer Gescgwindigkeitsverteilungen, Nachr. Wiss fachgruppe, Gottingen, math. Phys., 1, 79-114. Translated as, General instability criterion of laminar velocity disturbances, NACA TM-792, 1936.
4  C.-C. Lin, On the stability of two-dimensional parallel flows, Proc. NAS, 30, 1944, 316-324.
5  R. Fjørtoft, Application of integral theorems in deriving criteria of stability for laminar flows and for the baroclinic circular vortex. Geofys. Publ., 1950, 17: 1-52
6  C.-C. Lin, The Theory of Hydrodynamic Stability, Cambridge Press, Cambridge, 1955, 1-153 .
7  P. G. Drazin and W. H. Reid, Hydrodynamic stability, Cambridge University Press, 2$^{nd}$ Ed., Cambridge, England, 2004, 69-123.

8 P. J. Schmid and D.S. Henningson., Stability and transition in shear flows, New York, Springer-Verlag, 2000.
9 F. M. White, Viscous fluid flow, 2$^{nd}$ Edition, McGraw-Hill, New York, 1991, 335-337.
10 H. Schlichting, Boundary Layer Theory, Springer, 7th Ed., Berlin, 1979.
11 O. Reynolds, An experimental investigation of the circumstances which determine whether the motion of water shall be direct or sinuous, and of the law of resistance in parallel channels, Phil. Trans. Roy. Soc. London A, 174, 1883, 935-982.
12 J. Peixinho, T. Mullin, Finite-amplitude thresholds for transition in pipe flow, J. Fluid Mech. 582, 2007, 169-178.
13 H.-S. Dou, Mechanism of flow instability and transition to turbulence, International Journal of Non-Linear Mechanics, 41(2006), 512-517. http://arxiv.org/abs/nlin.CD/0501049
14 H.-S. Dou, Physics of flow instability and turbulent transition in shear flows, International Journal of Physical Science, 2011, accepted and in press. Also, Technical Report of National University of Singapore, 2006; http://arxiv.org/abs/physics/0607004. Also as part of the invited lecture: H.-S. Dou, Secret of Tornado, International Workshop on Geophysical Fluid Dynamics and Scalar Transport in the Tropics, NUS, Singapore, 13 Nov. - 8 Dec., 2006.
15 H.-S. Dou, B.C. Khoo, and K.S. Yeo, Instability of Taylor-Couette Flow between Concentric Rotating Cylinders, Inter. J. of Therm. Sci., 47, 2008, 1422-1435. http://arxiv.org/abs/physics/0502069.
16 H.-S. Dou, Viscous instability of inflectional velocity profile, Proceedings of the Forth International Conference on Fluid Mechanics, Ed. by F. Zhuang and J. Li, Tsinghua University Press & Springer-Verlag, July 20-23, 2004, Dalian, China, pp.76-79.
17 H.-S. Dou，Three important theorems for flow stability, Proceedings of the Fifth International Conference on Fluid Mechanics, Ed. by F. Zhuang and J. Li, Tsinghua University Press & Springer, 2007, pp.57-60. http://www.arxiv.org/abs/physics/0610082
18 H.-S. Dou, B.C. Khoo, Criteria of turbulent transition in parallel flows, Modern Physics Letters B, 24 (13), 2010, 1437-1440. http://arxiv.org/abs/0906.0417
19 M. Nishi, B. Unsal, F. Durst, G. Biswas, Laminar-to-turbulent transition of pipe flows through puffs and slugs, J. Fluid Mech., 614, 2008, 425-446.
20 B. Hof, C.W.H. van Doorne, J. Westerweel, F.T.M. Nieuwstadt, H. Faisst, B. Eckhardt, H. Wedin, R.R. Kerswell, F. Waleffe, Experimental observation of nonlinear traveling waves in turbulent pipe flow, Science, 305, No.5690, 2004, 1594-1598.
21 H. Shan, Z. Zhang, F.T.M. Nieuwstadt, Direct numerical simulation of transition in pipe flow under the influence of wall disturbances, Inter. J. of Heat and Fluid Flow, 19 (1998) 320-325.
22 X.J. Wang, J.S. Luo, H. Zhou, Inherent mechanism of "breakdown" process during laminar flow-turbulence transition in channel flows, Science in China Ser. G, 2005, 35(1): 71-78 (In Chinese).

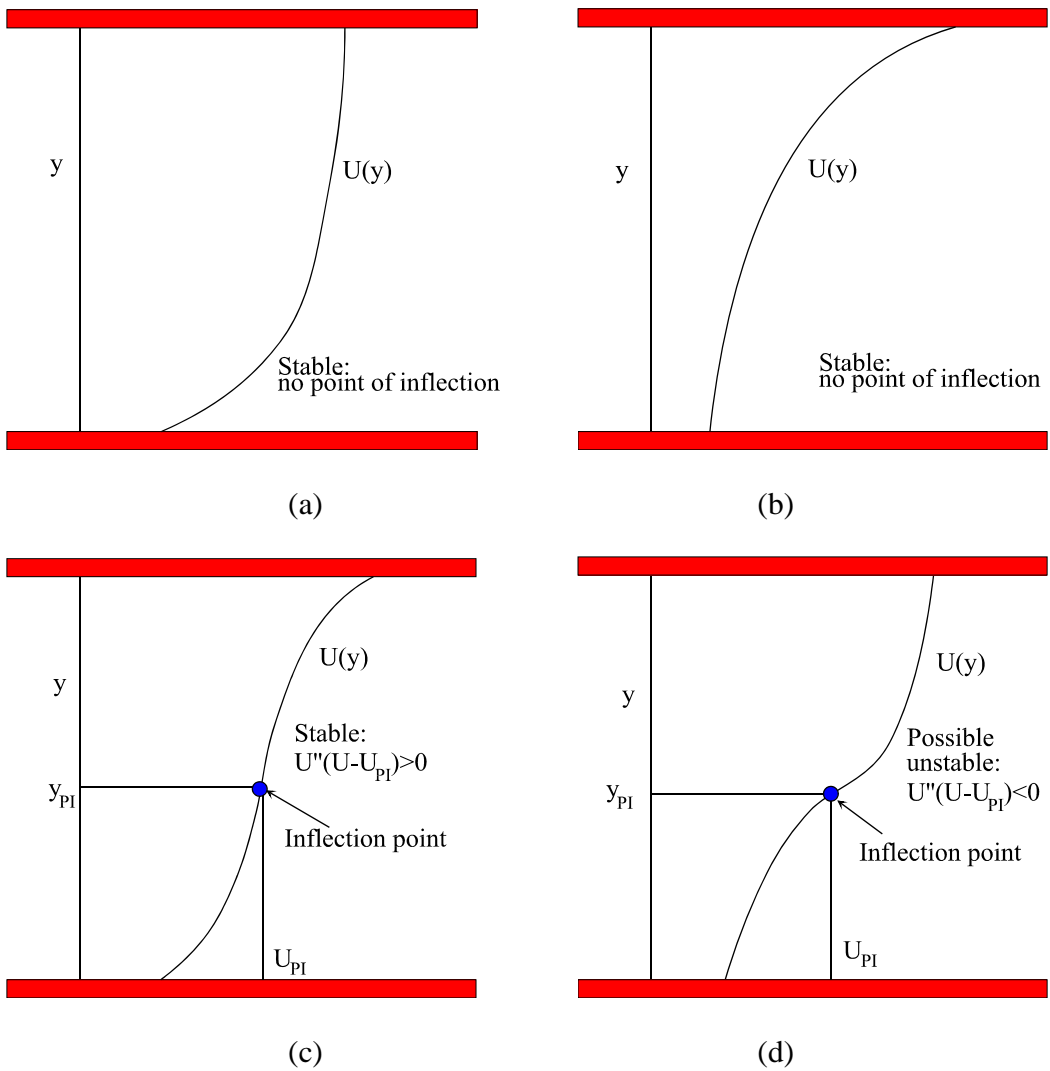

Fig.1 Four candidate inviscid velocity profiles evaluated from Rayleigh Theorem (1880) and Fjørtoft Theorem (1950) (adapted from White, 1991; and Drazin and Reid, 2004).

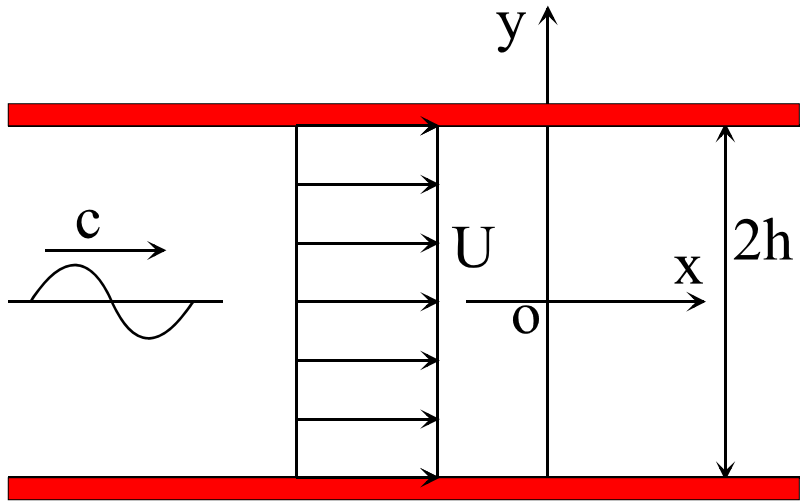

Fig.2 Inviscid parallel flow between two parallel walls; the base flow is a uniform flow.

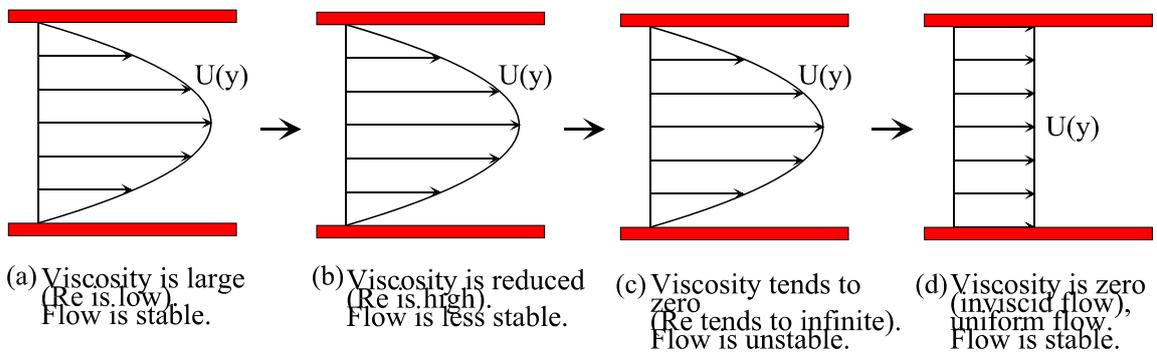

(a) Viscosity is large (Re is low). Flow is stable.
(b) Viscosity is reduced (Re is high). Flow is less stable.
(c) Viscosity tends to zero (Re tends to infinite). Flow is unstable.
(d) Viscosity is zero (inviscid flow), uniform flow. Flow is stable.

Fig.3 Evolution of stability bahaviour of velocity profile between two parallel walls with the variation of viscosity.